\documentstyle[epsfig,12pt]{article}
\newcommand{\mycomm}[1]{\hfill\break
$\phantom{a}$\kern-3.5em{\tt===$>$ \bf #1}\hfill\break}
\newcommand{\mycommA}[1]{\hfill\break
$\phantom{a}$\kern-3.5em{\tt***$>$ \bf #1}\hfill\break}

\catcode`\@=11 
\let\rel@x=\relax
\newcount\timecount
\newcount\hours \newcount\minutes  \newcount\temp \newcount\pmhours
\hours = \time
\divide\hours by 60
\temp = \hours
\multiply\temp by 60
\minutes = \time
\advance\minutes by -\temp
\def\hour{\the\hours}
\def\minute{\ifnum\minutes<10 0\the\minutes
            \else\the\minutes\fi}
\def\clock{
\ifnum\hours=0 12:\minute\ AM
\else\ifnum\hours<12 \hour:\minute\ AM
       \else\ifnum\hours=12 12:\minute\ PM
            \else\ifnum\hours>12
                 \pmhours=\hours
                 \advance\pmhours by -12
                 \the\pmhours:\minute\ PM
                 \fi
            \fi
         \fi
\fi
}

 \def\monthname{\rel@x\ifcase\month 0/\or January\or February\or
   March\or April\or May\or June\or July\or August\or September\or
   October\or November\or December\else\number\month/\fi}

\def\bold#1{\setbox0=\hbox{$#1$}     \kern-.025em\copy0\kern-\wd0
     \kern.05em\copy0\kern-\wd0
     \kern-.025em\raise.0433em\box0 }

\def\lsim{\mathrel{\mathpalette\@versim<}}
\def\gsim{\mathrel{\mathpalette\@versim>}}
\def\@versim#1#2{\vcenter{\offinterlineskip
        \ialign{$\m@th#1\hfil##\hfil$\crcr#2\crcr\sim\crcr } }}
\catcode`\@=12

\begin{document}
\def\beq{\begin{equation}}
\def\eeq{\end{equation}}
\def\MSbar {\hbox{$\overline{\hbox{MS}}\,$}}
\def\eff{\hbox{\it\footnotesize eff}}
\def\FP{\hbox{\it\footnotesize FP}}
\def\APT{\hbox{\it\footnotesize APT}}
\def\mysim{\kern -.1667em\lower0.8ex\hbox{$\tilde{\phantom{a}}$}}
\vskip 20pt

\begin{titlepage}
\begin{flushright}
{\footnotesize
TAUP-2503-98 \\}
\end{flushright}
\begin{centering}

{\large{\bf
Can the QCD running coupling have a causal analyticity structure?
}}
\vskip 30pt
\vskip 30pt
{\bf Einan Gardi $^a$} ,\,\,\, {\bf Georges Grunberg $^b$} \,\,\, and
\,\,\, {\bf Marek Karliner $^a$} \\

\vspace{.2in}

$^a$ School of Physics and Astronomy
\\ Raymond and Beverly Sackler Faculty of Exact Sciences
\\ Tel-Aviv University, 69978 Tel-Aviv, Israel
\\ e-mail: gardi@post.tau.ac.il, marek@proton.tau.ac.il

\vspace{.2in}

$^b$ Centre de Physique Th\'eorique de l'Ecole Polytechnique
\footnote{CNRS UMR C7644}
\\ 91128 Palaiseau Cedex, France
\\email: grunberg@cpht.polytechnique.fr

\vspace{0.9cm}

{\bf Abstract} \\
\vspace{0.35cm}
{\small
Solving the QCD renormalization group equation at the 2-loop and 3-loop
orders we obtain explicit expressions for the coupling as a function
of the scale in terms of the Lambert W function.
We study the nature of the ``Landau singularities'' in the complex
$Q^2$ plane and show that perturbative freezing can lead, in certain
cases, to an analyticity structure that is consistent with causality. 
We analyze the Analytic Perturbation Theory (APT) approach which is
intended to remove the ``Landau singularities'', and show
that at 2-loops it is uniquely defined in terms of the Lambert W
function, and that, depending on the value of the first two $\beta$
function coefficients $\beta_0$ and $\beta_1$, it is either consistent
with perturbative freezing (for $\beta_1<-\beta_0^2$) 
with an infrared limit of $-\beta_0/\beta_1$ or leads to a
non-perturbative infrared coupling with a limit of $1/\beta_0$ 
(for $\beta_1>-\beta_0^2$).
The possibility of a causal perturbative coupling is in accordance with
the idea that a purely perturbative Banks-Zaks phase with an infrared
fixed-point exists in QCD if the number of flavours ($N_f$) is increased.
The causality condition implies that the perturbative phase is realized
for $N_f\geq 10$. 
} 
\end{centering}

\end{titlepage}
\vfill\eject

\section{Introduction}

Due to asymptotic freedom \cite{oneloop}, physical quantities in QCD at
large momentum transfers $Q^2\equiv -q^2$, where $q^2$ is a space-like
momentum-squared, can be calculated as power expansions 
in the coupling constant $x(Q^2)=\alpha _s(Q^2)/\pi$. 
This seems as a reasonable expansion\footnote{Even though the series 
actually does not converge, and in-fact it is non Borel-summable.}  
since the coupling vanishes at this limit according to: 
\begin{equation}
x(Q^2)\sim \frac{1}{\beta_0 \ln \left( Q^2/\Lambda ^2\right) }
\label{1_loop_running}
\end{equation}
where 
\beq
\beta _0=\frac 14\left( 11-\frac 23N_f\right) 
\label{beta0}
\eeq
and $\Lambda $ is the QCD scale. 
$x(Q^2)$ in (\ref{1_loop_running}) is a solution to the
1-loop renormalization group (RG) equation 
\beq
Q^2\frac{dx}{dQ^2}=-\beta_0\,x^2.
\label{1loopRG}
\eeq
Going to lower $Q^2$, $x(Q^2)$ becomes larger, and higher loops 
have to be taken into account in (\ref{1loopRG}), as well as in
the expansions that describe physical quantities in terms of
$x(Q^2)$.
The RG equation at the $(n+1)$-th loop order is
\beq
Q^2\frac{dx}{dQ^2}=\beta(x)=-\beta_0\,x^2(1+cx+c_2x^2+\cdots c_nx^n).
\label{n1loopRG}
\eeq

For $Q^2$ below $\Lambda^2$, non-perturbative
effects, which are non-expandable in $x(Q^2)$, become dominant, and the
perturbative expansion becomes useless. However, in the intermediate
regime, the perturbative solution can be fitted to the data, provided
it is supplemented by power-like terms. These terms are
non-perturbative but they can be characterized by perturbation theory,
as they are related to the large order asymptotics of the perturbation
series, and in particular to renormalons (see \cite{Bridge} and refs. therein).
It is the non-convergence of the perturbative expansion, that provides
some information on the non-perturbative terms.

Another indication that the perturbative result is
incomplete, and cannot describe the low $Q^2$ physics unless it is
supplemented by non-perturbative corrections, comes from considering its
analyticity structure in the complex $Q^2$ plane:
a generic QCD observable, that depends on a space-like momentum
variable $Q^2$, is expected to be an analytic function of $Q^2$ in the
entire complex plane, except the negative real (time-like) axis. 
Singularities on the time-like axis are meaningful since they correspond 
to production of on-shell particles, while existence of singularities
away from the time-like axis would violate causality. 
Thus causality constrains the functional dependence of observables on 
$Q^2$. The analyticity condition is equivalent to the requirement that $x(Q^2)$
obeys the following dispersion relation:  
\beq
\beta_R(s)\equiv -\frac{1}{\pi}{\rm Im}\{ x(-s)\}
\label{beta_R}
\eeq
with
\beq
x(Q^2)=- \int_0^\infty ds\frac{\beta_R(s)}{s+Q^2}
\label{DR}
\eeq
where $\beta_R(s)$, with $0\leq s< \infty$, 
is called the spectral density function.  
 
A 1-loop perturbative result for a generic QCD observable depends on
the coupling (\ref{1_loop_running}) and therefore 
contains a ``Landau-pole'' at $Q^2=\Lambda^2$.
This pole is located on the positive real axis (the space-like axis)
and therefore it is non-physical. This pole is expected to be washed
out when further perturbative and non-perturbative corrections are added.  
In general, higher loop perturbative results for the
coupling have a more complicated analyticity structure which is still
inconsistent with causality. Thus, in general, causality can by
saved only be inclusion of non-perturbative terms. 

A special case in which the perturbative 
result by itself can be consistent with causality, is
when the perturbative $\beta$ function (\ref{n1loopRG}) has a
zero. Then, the coupling reaches a finite value in the infrared limit
and thus it is finite for any real and positive $Q^2$.
To be consistent with causality, the coupling should be non-singular 
in the entire $Q^2$ plane except the time-like axis, and therefore
there may be cases where freezing occurs but causality is still
violated, due to complex $Q^2$ singularities. 

In real-world QCD, with only three light flavors, 
it seems from the first few terms in the $\beta$ function 
series that there is no perturbative freezing \cite{FP}.  
In this case, the standard perturbative approach always faces
the ``Landau singularity'' problem. Refs. \cite{Analytic_1,Analytic}
revive a natural solution to this problem through the so-called  Analytic 
Perturbation Theory (APT) approach. 
The APT coupling  was used in \cite{dispersive_grunberg} for discussing power
corrections within the more general (non-perturbative) dispersive approach 
of \cite{dispersive}, and was
also considered there (as well as in refs. \cite{Bridge,Zakharov}) 
as a possible model for non-OPE power terms.

In this paper we solve the 2-loop and 3-loop RG equation giving 
explicit expressions for the coupling as a function
of the scale in terms of the Lambert W function.
This enables us to study the location and the nature of the 
``Landau singularities'' in the complex $Q^2$ plane. 
In particular we find exact criteria which determine when perturbative
freezing \`{a} la Banks-Zaks \cite{BZ} leads to 
an analyticity structure that is consistent with causality. 
Using the Lambert W solution, we then analyze the APT approach and 
show that it is either consistent
with perturbative freezing (for $c<-\beta_0<0$) 
with an infrared limit of $-1/c$, or leads to a
non-perturbative coupling with an infrared limit of $1/\beta_0$ 
(for $c>-\beta_0$). 

\section{The Lambert W function -- exact explicit coupling at 2-loops}

We start with a 2-loop RG equation 
\beq
\beta(x)=\frac{dx}{dt}=-\beta_0{x}^2\left(1+cx \right)
\label{beta_2loop}
\eeq
where $t=\ln(Q^2/\Lambda^2)$. 
Note that the value of $\Lambda$ here 
should be different from the one in
(\ref{1_loop_running}), so that the same phenomenological
value for $x(Q^2)$ will be obtained. 
When higher-order corrections are added, $\Lambda$
should be further adjusted, but for simplicity we use the same
notation throughout. The 2-loop coefficient 
$c$, which is renormalization scheme
independent, is given by:
\beq
c=\frac{\beta _1}{\beta _0}=\frac 1{4\beta _0}\left[ 102-\frac{38}
3N_f\right]   
\label{c}
\eeq
A straightforward integration of (\ref{beta_2loop}) for the
asymptotically-free case ($\beta_0>0$) yields:
\beq
\beta_0 \ln(Q^2/\Lambda^2)=\frac{1}{x}
-c\ln\left[\frac{1}{x}+c\right]
\label{2loop_int}
\eeq
At this stage it is already clear that $x(Q^2)$ has a cut on the
negative real axis, being a function of
$\ln(Q^2/\Lambda^2)$. The problem of inverting (\ref{2loop_int}) makes
it difficult to study the singularity structure. In
\cite{Analytic_1}, for instance, the inversion of (\ref{2loop_int})
relies on the assumption that the $1/x$ term is much larger than 
the logarithmic term. 
Although correct in the deep perturbative region, this
approximation is inadequate for studying the singularity structure of 
$x(Q^2)$. In particular, this approximation does not allow
perturbative freezing at all, since in the latter case the logarithmic term
becomes dominant in the infrared region. 
 
Luckily, an explicit solution of the 2-loop RG equation (\ref{beta_2loop}),
i.e. an inversion of (\ref{2loop_int}), can be written in terms of the
so-called Lambert $W$ function \cite{Lambert}. $W(z)$ is defined by
\beq
W(z) \exp\left[W(z)\right]=z
\label{LambertW_def}
\eeq
The coupling is then given by:
\beq
\begin{array}{c}
\displaystyle
x(Q^2)=-\frac{1}{c}\,\,\frac{1}{1+W(z)} \nonumber\\
\phantom{a}\\
\displaystyle
z = -\frac{1}{c}\exp\left(-1-\beta_0 t/c\right)
=
-\frac{1}{c\, e}
\left(\frac{Q^2}{\Lambda^2}\right)^{-\beta_0/c}
\end{array}
 \label{W_sol_2loop}
\eeq
$W(z)$ is a multi-valued function with an infinite number of branches,
denoted by $W_n(z)$. We follow \cite{Lambert} as for the division of
the branches and notation. The requirement that $x(Q^2)$ is
real and positive for a real positive $Q^2$ (at least for $Q^2\gg \Lambda^2$),
is sufficient to determine the physical branch depending on the sign
of $c$:
\begin{description}
\item{(a) } for $c>0$, $z<0$ and the physical branch is $W_{-1}(z)$. 
This branch is a real monotonically decreasing function for 
$z\in(-1/e,0)$, with $W_{-1}(z)\in(-\infty,-1)$ (outside this range
$W_{-1}(z)$ is complex).
Thus, the ultraviolet limit corresponds to $z\longrightarrow 0^-$, 
$W_{-1}(z)\longrightarrow -\infty$, and $x\longrightarrow 0^+$, as
required by asymptotic freedom. In the infrared region, below the
``Landau singularity'' (see further discussion later) $x$ is
complex. 
\item{(b) } for $c<0$, $z>0$ and the physical branch is the principal
  branch, $W_0(z)$. This branch is a real monotonically
increasing function for $z\in(-1/e,\infty)$, with
$W_0\in (-1,\infty)$ (outside this range,
$W_0(z)$ is complex). For $z\geq0$, $W_0(z)\in [0,\infty)$.
The ultraviolet limit corresponds to $z\longrightarrow\infty$,
$W_0(z)\longrightarrow\infty$ and $x\longrightarrow 0^+$. The
infrared limit corresponds to $z\longrightarrow 0^+$,
$W_0(z) \longrightarrow 0^+$, and $x\longrightarrow (-1/c)^-$ which is
consistent with the Banks-Zaks perturbative fixed-point value
$x_{FP}=-1/c$. 
\end{description}

The solution in (\ref{W_sol_2loop}) can be quite useful whenever a
two-loop evolution of the QCD coupling is required. It clearly yields
more accurate results than the standard expansion of the coupling in
$1/\ln(Q^2/\Lambda^2)$. Note that the latter can be obtained from
(\ref{W_sol_2loop}) using the asymptotic formulae for $W(z)$:
for $c<0$ one uses the asymptotic expansion of $W_0(z)$ at large
positive $z$ that starts with $W_0(z)\sim \ln(z)-\ln(\ln(z))$, 
and for $c>0$ one uses the asymptotic expansion of $W_{-1}(z)$ at small
negative $z$ that starts with $W_{-1}(z)\sim \ln(-z)-\ln(-\ln(-z))$.

Other possible 
applications of (\ref{W_sol_2loop}) are in the context of resummation
methods, such as the iterated construction suggested by Maxwell in 
\cite{Maxwell} and the improved Baker-Gammel approximants suggested by
Cveti\v{c} in \cite{Cvetic} which both generalize the diagonal Pad\'e 
approximants approach \cite{diag_PA}.

The general idea in these resummation methods is that knowledge of the 
first few orders in a perturbative expansion, together with the RG
equation, can be used to construct approximants to physical quantities
that have a better accuracy than the truncated perturbative
series, and exhibit reduced renormalization scale and scheme
dependence (for related ideas, see \cite{ECH,PMS}).
Diagonal Pad\'e Approximants, as opposed to truncated series, 
are exactly invariant with respect to an arbitrary renormalization 
scale transformation, so
long as the transformation respects the 1-loop evolution form (i.e. an
evolution related to the 1-loop $\beta$ function) \cite{diag_PA}. 
In order to achieve scale invariance beyond this
approximation, it turns out \cite{Maxwell,Cvetic} that one has to use more
complicated functions, which are related to the inverted solution of the 
some higher-order RG equation. Using the explicit
inverted solution of the 2-loop RG equation (\ref{W_sol_2loop}) the
methods of \cite{Maxwell,Cvetic} can be easily implemented in practice
and also their mathematical structure can be further studied.

It is worth mentioning that another way\footnote{An exact solution of 
(\ref{2loop_int}) in term of a (modified) Borel representation can
also be found in \cite{grunberg-borel}.} to invert (\ref{2loop_int})
was suggested in the past by Khuri and Ren \cite{Khuri}: they wrote
the  2-loop coupling in terms of $F(\zeta)$, which is defined by
$2F(\zeta)-\exp[F(\zeta)]+1=\zeta$. 
The latter equation was considered earlier in
different physical contexts (see refs. in \cite{Khuri}). 
There is, of course, a one-to-one correspondence between the 
Lambert W solution of (\ref{W_sol_2loop}) and that of \cite{Khuri}:
\beq
-\frac12\exp{\left[F(\zeta)\right]}
=W\left(-\frac12 \exp\left[\frac12(\zeta-1)\right]\right).
\eeq 
In the following we stick with the Lambert W solution.

\section{The singularity structure of the 2-loop coupling}

Having found an explicit solution for coupling (\ref{W_sol_2loop}), we
would like now to analyze its singularity structure in the complex $Q^2$ 
plane and, in particular, to find when it is consistent with causality. 
To do so it is essential to define the analytical continuation of $x(Q^2)$ 
to the whole complex $Q^2$ plane, namely to
specify the branch of $W(z)$ that is used in 
(\ref{W_sol_2loop}) for 
a generic $Q^2=\vert Q^2\vert e^{i\phi}$, where $-\pi<\phi<\pi$.

At this stage it is necessary to give a brief description of the 
singularity structure of the Lambert W function; more details can be
found in \cite{Lambert}. The partition of the complex $W$ plane 
between the different branches, $W_0$, $W_{\pm 1}$ and $W_{\pm 2}$, 
is shown by dashed lines in fig.~1 (the other lines in fig.~1 will be
discussed later).   
It is important to realize that the singularity structure of the
different branches $W_n(z)$ is different. 
$W_0$, $W_{-1}$ and $W_1$ share a common branch point
at $z=-1/e$, at which $W=-1$. The $z=-1/e$ cut in all three
branches is chosen to be $z\in (-\infty,-1/e)$. 
While for $W_0$, $z=-1/e$ is
the only singularity, other branches $W_n(z)$ with $n\neq 0$ have a
branch point at $z=0$ with a cut at $(-\infty,0)$. The
$z=0$ cut is the only singularity in $W_n$ for $n\geq2$.
Note that $W(y)$ obeys the following symmetry \cite{Lambert},
\beq
W^*_{-n}(y^*)=W_n(y)
\label{summetry}
\eeq
and therefore it is 
possible to obtain $W(z)$ and $x(Q^2)$ on the upper half-plane from
those on the lower half-plane. 
Note also that the branches $W_{-1}(z)$ and $W_{1}(z)$ have a
common real limit along the cut for $z\in (-1/e,0)$. It is only a matter of
convention that the negative real axis ($W\in (-\infty,-1/e)$) belongs
to $W_{-1}$ rather than to $W_1$.      

The criterion we use for defining the analytical continuation is that for
$\vert Q^2\vert \gg \Lambda^2$, the coupling $x(Q^2)$, and therefore
also $W(z)$, will be continuous as a function of the phase of $Q^2$. 
From the analyticity structure of
$W(z)$ described above it is clear that as long as the phase of $z$
does not reach $\pm \pi$ for any $Q^2$ in the first sheet, the only
relevant branch is the physical one, i.e. the one that yields
a real positive $x$ for real positive $Q^2$: $W_0$ for $c < 0$ and
$W_{\pm 1}$ for $c > 0$.
Let us define $z=\vert z\vert e^{i\delta}$.   
Using (\ref{W_sol_2loop}) the condition $-\pi <\delta<\pi$ 
can be directly translated to conditions on $\beta_0$ and $c$, and
therefore further division of the $c$ axis is suggested as follows:
\begin{description}
\item{(a) } for $c<-\beta_0<0$ we have 
$\delta=-(\beta_0/c)\phi$, and for $\phi \in (-\pi,\pi)$ we always have 
$-\pi <-\vert\beta_0/c\vert\pi<\delta<\vert\beta_0/c\vert\pi<\pi$. 
Thus $W_0(z)$ is the only relevant branch, its boundary is never reached 
and no singularity is encountered. 
We conclude that here the coupling has a perturbative fixed-point at
$x_{\FP}=-1/c$, and an analyticity structure that is consistent with causality.
\item{(b) } for $-\beta_0<c<0$, we find that 
  $\delta=-(\beta_0/c) \phi$ reaches $\pm \pi$ at $\pm\phi_0$, with 
\beq
\phi_0\equiv \vert c/\beta_0\vert\, \pi. 
\label{phi0}
\eeq
Thus, when $Q^2$ has a phase of $\pm \phi_0$, the boundary 
of the $W_0(z)$ branch is reached. Consequently, an image of the cut 
$z\in(-\infty,-1/e)$ appears on the first sheet in the $Q^2$ plane.
The branch point corresponding to $z=-1/e$ appears at
\beq
Q^2=Q_0^2\,e^{\pm i\phi_0}
\label{complex_poles}
\eeq
where
\beq 
Q_0^2=\Lambda^2 \vert c\vert^{-c/\beta_0}.
\label{Q0}
\eeq
The cuts in the $Q^2$ plane and the analytical continuation for
$\vert \phi\vert>\phi_0$ will be discussed later.
\item{(c) } for $c>\beta_0/2$, the solution is given by $W_{-1}(z)$
  with $\delta=+\pi-(\beta_0/c)\phi$ for $\phi>0$ and by $W_{1}(z)$
  with $\delta=-\pi-(\beta_0/c)\phi$ for $\phi<0$. 
For $\phi\in(0,\pi)$, $\delta>(1-\beta_0/c)\pi$ and therefore it never
reaches the $W_{-1}$ branch boundary at $\delta=-\pi$. Similarly, for
 $\phi\in(-\pi,0)$, $\delta<(-1+\beta_0/c)\pi$ and therefore it never
reaches the $W_{1}$ branch boundary at $\delta=+\pi$.
Thus, the only relevant branches are $W_{\pm1}$ and the only singularity
  encountered is the one at $z=-1/e$ with the cut $z\in (-\infty,-1/e)$ 
that separates between $W_1$ and $W_{-1}$. 
From (\ref{W_sol_2loop}) we find that the $z=-1/e$
singularity appears on the positive real $Q^2$ axis, at $Q^2=Q_0^2$,
where $Q_0^2$ is given in (\ref{Q0}). 
The cut $z\in (-\infty,-1/e)$ corresponds to a cut on the infrared
section of the positive real $Q^2$ axis, $Q^2\in(0,Q_0^2)$. 
\item{(d) } for $0<c<\beta_0/2$, $\delta=\pm\pi-(\beta_0/c)\,\phi$ and the
boundary of $W_{-1}$ at $\delta=-\pi$, and that of $W_{1}$ at $\delta=\pi$,
is reached at $\phi=\pm\phi_1$, where $\phi_1= 2\,(c/\beta_0)\, \pi$. 
Like in case (b) above, it is required to define $W(z)$ for
$\phi>\phi_1$. This will be done soon.
\end{description}    
 
In the two cases with large $\vert\beta_0/c\vert$, (b) and (d) above, 
we found that the
boundary of the physical branch is reached, and therefore a definition
of $W(z)$ in eq. (\ref{W_sol_2loop}) for $\vert \phi\vert >\phi_{0}$,
in (b), and $\vert \phi\vert >\phi_{1}$, in (d), is required.
The criterion that $W(z)$ should be continuous as a function of $\phi$
for large $\vert Q^2\vert$ implies a particular definition for the analytical 
continuation of $W(z)$: starting with a given branch $W_{\vert n\vert} (z)$
and increasing $\vert \phi\vert $, then when the branch boundary is 
reached one switches to the next branch $W_{\vert n\vert+1}(z)$.  

Let us illustrate the above starting with the case $0<c<\beta_0/2$, 
(d) above, and consider a
gradual increase in $\vert \phi\vert $, the phase of $Q^2$, in the
lower half-plane $\phi<0$. For a positive real
$Q^2$, $z<0$ and $W(z)$ is just on the boundary between $W_{-1}$ and
$W_{1}$, shown by the dashed line in fig.~1. 
For a small negative $\phi$, $z$ is slightly below the negative real
axis ($\delta=-\pi-(\beta_0/c)\phi$) and the solution is within the branch
$W_1(z)$. This is exemplified in fig.~1 by the case $g=0.1$ ($\phi=-0.1\pi$).
When $\phi=-\phi_1=-2\,(c/\beta_0)\,\pi$, $\delta=\pi$, 
$z$ becomes negative real
again, and so the cut $z\in (-\infty,0)$ which is the upper boundary of the
$W_1(z)$ branch is reached. Continuity of $W(z)$ is obtained if, and
only if $W_2(z)$ is used for $\phi< -\phi_1$. In fig.~1, this is
exemplified by the case $g=0.6$ ($\phi=-0.6\pi$). 
Note that in the specific example
given in fig.~1, the time-like axis $\phi=-\pi$ is reached within
the $W_2$ branch (the dot-dash line), but in general, depending on the
ratio $\beta_0/c$, higher branches of $W$ may become relevant.

Next, consider the case $-\beta_0<c<0$, (b) above, where the positive
real $Q^2$ solution is the positive real $W$ axis in the $W_0$ branch,
described by the $g=0$ line in fig.~2. For a small negative
$\phi$, $\delta=-(\beta_0/c) \phi$ is still far from the $\delta=-\pi$
limit and the solution is
given by $W_0(z)$. In fig.~2, this is exemplified by the case $g=0.1$ 
($\phi=-0.1\pi$).
For $\phi=-\phi_0=\,(c/\beta_0)\,\pi$, the cut at $z=(-\infty,-1/e)$
which is the lower boundary of the $W_0$ branch is reached. Continuity
of $W(z)$ at large $\vert Q^2\vert$ (the right side of fig.~2) implies
that for $\phi\leq-\phi_0$, the branch $W_{-1}(z)$ should be
used. In fig.~2 this is exemplified by the case $g=0.7$
($\phi=-0.7\pi$). 
The possibility that higher branches ($W_{\vert n\vert} $ for $n>1$) 
will become relevant (depending on the ratio $\beta_0/c$) exists also
here. Just like in the previous case, these branches are reached
through the $z=0$ cut.
   
A unique feature of the analytical continuation we perform appears 
in case (b) (see fig.~2): this is a ``phase transition'' from the
Banks-Zaks non-trivial infrared fixed-point to the trivial one, that
occurs at $\phi=\pm\phi_0$. For any $\vert \phi\vert <\phi_0$ (the
lines with $g\leq 0.6$ in fig.~2) $W(z)$ flows to zero in the infrared,
which corresponds to the perturbative fixed-point at $x_{\FP}=-1/c$. 
On the other hand, for any $\vert \phi\vert >\phi_0$ (the lines with 
$g\geq0.7$ in fig.~2) $\vert W(z)\vert$ becomes infinite in the infrared, which
corresponds to $x(Q^2)\longrightarrow 0$. The boundary between the $W_0$
branch and the $W_{-1}$ branch in fig.~2 separates between the two regimes. A
flow from the ultraviolet down to the infrared for $\phi=\phi_0$
($\phi=-(2/3)\pi$ in fig.~2)
leads to a fork at $W=-1$ (corresponding to $Q^2=Q_0^2$) from which
there are two ways to continue towards $Q^2=0$, one is to the right,
within the $W_0$ branch leading to $W=0$, and the other to the left, 
on the boundary between the $W_{-1}$ and $W_1$ branches, 
leading to $W=-\infty$. In other words, choosing the analytical continuation
beyond $\vert\phi\vert=\phi_0$ as we did, guarantees continuity for 
$Q^2\in(Q_0^2,\infty)$, but leaves a discontinuity in the complex
$Q^2$ plane, along the $\phi=\pm\phi_0$ directions for any
$Q^2\in(0,Q_0^2)$.

Let us now summarize how we choose the branch of the Lambert W
function in (\ref{W_sol_2loop}) according to the proposed analytical 
continuation of the coupling to the entire complex $Q^2$ plane.
It is convenient to determine the branch from $d\equiv -(\beta_0/c)\,\phi$,
where $\phi$ is the phase of $Q^2$, as before\footnote{Note that $d$
is not the phase of $z$, since the latter has an additional $\pi$ term
for $c>0$.}.
Given $d$, we use the branch $W_n$ such that for $c<0$,
\beq
d\in((2n-1)\,\pi\,,\,(2n+1)\,\pi]
\label{d_n_neg_c}
\eeq
and for $c>0$,
\begin{eqnarray}
\label{d_n_pos_c}
d\in(2(n-1)\,\pi\,,\,2n\pi]      & n\geq 1   \\ \nonumber
d\in(2n\pi\,,\,2(n+1)\pi]        & n\leq -1
\end{eqnarray}
For instance, for $c<0$, $W_{-1}$ is used for
$d\in(-3\pi,-\pi]$ and $W_0$ is used for $d\in(-\pi,\pi]$, as shown in 
fig.~2. 
For $c>0$, $W_{1}$ is used for $d\in(0,2\pi]$ and $W_2$ is used for 
$d\in(2\pi,4\pi]$, as shown in fig.~1.

Note that the lower half of the complex $Q^2$ plane ($\phi<0$) 
always corresponds to a positive imaginary part for the coupling. 
For $c<0$ the lower half-plane corresponds to $d<0$
which, according to (\ref{d_n_neg_c}), refers to $n\leq 0$ i.e. 
${\rm Im}\{W\}<0$ that yields ${\rm Im}\{x(Q^2)\}>0$ in (\ref{W_sol_2loop}). 
For $c>0$, $\phi<0$ corresponds to $d>0$ which according to 
(\ref{d_n_pos_c}) refers to $n\geq 1$ i.e. ${\rm Im}\{W\}>0$ that 
again yields ${\rm Im}\{x(Q^2)\}>0$ in (\ref{W_sol_2loop}).

We are now in a position to address
the question we started with, namely what is the
singularity structure of $x(Q^2)$ in the complex $Q^2$ plane.
There are three different possibilities,
depending on the values of $c$ and $\beta_0>0$, 
which are shown in fig.~3\footnote{Some of these conclusions have been
  anticipated by Uraltsev \cite{Uraltsev_private}.}:
\begin{description}
\item{(a) } for $c<-\beta_0<0$ the 2-loop coupling has a non-trivial 
  perturbative infrared
  fixed point at $x_{\FP}=-1/c$ and an analyticity structure that is
  consistent with causality. Note there is no pole in the denominator of 
  (\ref{W_sol_2loop}), since whenever $W$ is real, it is positive.
 From considering only the 2-loop $\beta$ function it is not possible
 to exclude the physical relevance of this fixed-point,
 and there is no indication of the existence of non-perturbative effects.     
\item{(b) } for $-\beta_0<c<0$ the $\beta$ function seems to have a
  fixed-point, as in a), but since $x_{\FP}=-1/c$ is large here, 
this fixed-point is probably not reliable -- it can be 
  washed out by higher-loop contributions or by non-perturbative effects. 
Existence of the latter is strongly suggested by the presence of 
causality violating singularities: there are two branch points at 
$Q^2=Q_0^2e^{\pm i\phi_0}$ with radially oriented cuts that end at $Q^2=0$. 
The pole in the denominator of (\ref{W_sol_2loop}) at $W=-1$ falls
right on top of these branch points.
\item{(c) } for $c>0$ there is no infrared fixed-point. The singularity
  structure violates causality, due to the branch point on the
  space-like axis at $Q^2=Q_0^2$. 
 The pole in the denominator of (\ref{W_sol_2loop}) at $W=-1$ falls
right on top of this branch point. Note that for $c>0$ it is not
  important whether $\beta_0/c$ is large or small: in any case the
  only singularity is the single image of the $z=-1/e$ branch point, the
  starting point of the cut that separates between $W_1$ and $W_{-1}$ 
(There are no similar singularities for higher branches 
$\vert n\vert\geq 2$ of the W function). The non-physical cut on
the space-like axis is probably removed from any physical quantity by 
non-perturbative effects.

\end{description}

The truncated 2-loop $\beta$ function (\ref{beta_2loop}) can be 
considered as a
legitimate choice of renormalization scheme -- the so-called `t Hooft scheme.
However, in certain cases this may be a peculiar choice of scheme for
calculating observable quantities: the truncation of the $\beta$ 
function at the 2-loop order seems quite arbitrary. Therefore, it is
interesting to see to what extent our solution depends on this
specific choice of scheme.

\section{The Lambert W solution at 3-loops}

Given the above motivation we would like 
to generalize our results by considering a generic 3-loop RG equation:
\beq
\beta(x)=\frac{dx}{dt}=-\beta_0{x}^2\left(1+cx +c_2x^2\right)
\label{beta_3loop}
\eeq
where again $t=\ln(Q^2/\Lambda^2)$.   
A straightforward integration of (\ref{beta_3loop}) yields a function
that involves arctanh terms in addition to the terms of
the form appearing in (\ref{2loop_int}). Therefore, it is even harder
to explicitly invert this function.
To avoid this difficulty,  we use the following trick: we alter the
3-loop $\beta$ function slightly, by taking its $x^2$[1/1] Pad\'e
Approximant (PA)\footnote{PA's were found useful in various 
  applications in perturbative QCD -- see refs. \cite{diag_PA,PA_QCD}.}:
\beq
\beta_{PA}(x)=-\beta_0x^2\frac{1+[c-(c_2/c)]x}{1-(c_2/c)x}
\label{PA}
\eeq
This change does not limit generality as long as we restrict our
interest to 3-loops, since the difference between the $\beta$ function
in (\ref{PA}) and that in (\ref{beta_3loop}) is of the order ${\cal
  O}(x^4)$. Indeed, using (\ref{PA}) instead of (\ref{beta_3loop}) is
expected to change the singularity structure of $x(Q^2)$, but we
insist that (\ref{PA}) corresponds to a choice of scheme at 4-loops
and beyond which is just as arbitrary as the truncation in
(\ref{beta_3loop}).   Using (\ref{PA}) we obtain:
\beq
\beta_0 \ln(Q^2/\Lambda^2)=\frac{1}{x}
-c\ln\left[\frac{1}{x}+c-\frac{c_2}{c}\right]
\label{3loop_int}
\eeq
and finally, 
\beq
\begin{array}{c}
\displaystyle
x(Q^2)=-\frac{1}{c}\,\,\frac{1}{1-(c_2/c^2)+W(z)} \nonumber\\
\phantom{a}\\
\displaystyle
z = -\frac{1}{c}\exp\left[-1+(c_2/c^2)-\beta_0 t/c\right].
\end{array}
 \label{W_sol_3loop}
\eeq

Just like in the 2-loop case (\ref{W_sol_2loop}), the sign of $z$ and
therefore also the physically relevant branch of $W(z)$ are determined
according the sign of $c$: for $c>0$, $z<0$ and the physical branch is
$W_{-1}(z)$, taking real values in the
range $(-\infty,-1)$, while for $c<0$, $z>0$ and the physical branch
is $W_0(z)$, taking real values in the range $(0,\infty)$. 

Note, that the only significant difference between the 3-loop solution 
(\ref{W_sol_3loop}) and the 2-loop solution (\ref{W_sol_2loop}) is in
the relation between $W(z)$ and $x(Q^2)$. This is because the
difference in the definition of $z$ can be swallowed in a redefinition
of the scale parameter
\beq
\Lambda^2\longrightarrow\tilde{\Lambda}^2=\Lambda^2e^{c_2/(\beta_0 c)}.
\label{scheme_trans}
\eeq 
Consequently, we immediately know almost everything concerning the 
singularity structure of the PA-improved 3-loop coupling: 
it has the same type of
branch points and cuts that are described in fig.~3. 
The difference is, however, that the simple pole due 
to a zero in the denominator of $x(Q^2)$ in
(\ref{W_sol_3loop}) at $W(z)=c_2/c^2-1$, is not obtained on top of the
branch point(s) at $W=-1$ (which corresponds instead to the pole of 
the PA improved $\beta$ function).
Let us consider the three possible cases:
\begin{description} 
\item{(a) }
If $c>0$ then when $W$ is real ($W=W_{-1}$) it obeys $W(z)<-1$
and then any $c_2<0$ would result in a pole on the space-like axis. 
\item{(b) }
If $c<-\beta_0$ then when $W$ is real ($W=W_0$) it obeys $W(z)>0$ and
then if $c_2>c^2$ the pole will appear on the space-like axis
(for $c_2<c^2$ the coupling is causal) .
\item{(c) }
If $-\beta_0<c<0$, $W$ obtains (or approaches) any real value:
$W=W_0\in(0,\infty)$ on the space-like axis, $W\longrightarrow W_0\in(-1,0)$
for $\vert\phi\vert\longrightarrow \phi_0^-$ and 
$W\longrightarrow W_{\pm 1}\in(-\infty,-1)$ for 
$\vert\phi\vert\longrightarrow \phi_0^+$. Therefore for
$-\beta_0<c<0$, the pole will always appear, either on the
space-like axis (if $c_2>c^2$), or at $\vert\phi\vert=\phi_0$ (if $c_2<c^2$).
\end{description}

We conclude that the singularity structure of the
PA-improved 3-loop coupling is not much different from the 
2-loop coupling: there are regions of the parameter space
for which perturbative freezing is consistent with causality. 
In particular, this is true for small enough $\beta_0$ and $c<0$, 
which is the starting point for the Banks-Zaks expansion.
In general, however, higher-order terms in the $\beta$ function
create a more complicated singularity structure in the infrared
region, which is inconsistent with the analyticity requirement.

\section{The Analytic Perturbation Theory approach}

Recently it was suggested \cite{Analytic_1} to construct a causal
coupling constant by analytically continuing the coupling 
to the time-like axis (i.e. looking at $x(Q^2=-s)$ with $s>0$), use
the discontinuity, or equivalently the imaginary part, to define
a spectral function $\beta_R(s)$ as in (\ref{beta_R}) and finally use 
the dispersive integral (\ref{DR}) to construct a new space-like 
coupling $x_{\APT}(Q^2)$. APT stands for ``Analytic Perturbation
Theory'' since in this approach $x_{\APT}(Q^2)$ has the required
analyticity structure by construction, being defined through the
dispersion relation (\ref{DR}).

The authors of refs. \cite{Analytic_1,Analytic} show that the APT
coupling has some remarkable features such as a universal infrared 
limit of $1/\beta_0$ which is almost independent of  
the renormalization scheme and of the order of the $\beta$ function 
one starts with.  
A general argument for the universality of the APT infrared limit (for
$c>0$) was also given in \cite{dispersive_grunberg}.
With the explicit expression for the 2-loop coupling in
(\ref{beta_2loop}), we can now directly check these conclusions.

Before dealing with the 2-loop case, let us briefly review the APT
results at 1-loop, where both the spectral function and $x_{\APT}$ 
can be easily obtained in a closed form.
One starts with the 1-loop $\beta$ function (\ref{1_loop_running}) 
and performs an analytical continuation by substituting $Q^2=-s-i\epsilon$
where $s>0$ and $\epsilon>0$ is small. In the limit
$\epsilon\longrightarrow 0$ one obtains,  
\begin{eqnarray}
\label{Re_and_Im}
{\rm Im}\{x(-s)\}&=&\frac{\beta_0\pi}{(\beta_0\pi)^2+(\beta_0t)^2} \nonumber \\
{\rm Re}\{x(-s)\}&=&\frac{\beta_0t}{(\beta_0\pi)^2+(\beta_0t)^2}
\end{eqnarray}
where $t\equiv \ln(s/\Lambda^2)$. 
The spectral density is  
\beq
\beta_R(t)=-\frac{1}{\pi}{\rm Im}\{x(-s)\}=
-\frac{1}{\beta_0\left[\left(\ln(s/\Lambda^2)\right)^2+\pi^2\right]}.
\label{beta_R_APT}
\eeq
Finally, one constructs the corresponding space-like
effective coupling, through the dispersion relation (\ref{DR}). 
The integral can be performed analytically, and yields
\beq
x_{\APT}(Q^2)=\frac{1}{\beta_0}
\left[\frac{1}{\ln\left(Q^2/\Lambda^2\right)}
+\frac{\Lambda^2}
{\Lambda^2-Q^2}\right].
\label{removing_Landau}
\eeq
The first term in (\ref{removing_Landau})
is just the 1-loop perturbative result
and the second term exactly cancels the ``Landau-pole''. Since the second
term is a power correction, it does not alter the ultraviolet
behavior.
By construction, $x_{\APT}(Q^2)$ has a cut at ${\rm Re}\{ Q^2\}<0$ and
no other singularities in the complex plane, and is therefore
consistent with causality. The coupling has a non-perturbative
infrared fixed-point at $x_{\APT}(0)=1/\beta_0$.

It is clearly of interest to see how these APT results change at 2-loops.
Since we already know how to
analytically continue the space-like coupling to the entire
$Q^2$ plane, we can now examine $x(Q^2)$
on the time-like axis. Before doing so, we stress once more that for the
uniqueness of the APT solution it is necessary to impose a
continuity condition on  $x(Q^2)$ for large $\vert Q^2 \vert$.
This condition is implemented by
starting from $x(Q^2)$ on the space-like axis with $Q^2\gg\Lambda^2$,
and demanding continuity of $x(Q^2)$
as the phase  of $Q^2$ is changed.
This prescription is essential, since
the naive alternative of ``taking  a shortcut" and going directly
to the time-like axis
by trying to invert (\ref{2loop_int}) for negative $Q^2$
leads to an ambiguity in the choice of
the branch of $W$ and in the corresponding APT spectral function.

Actually, perturbatively speaking, there can be no ambiguity in the definition
of $x(Q^2)$ at complex $Q^2$.
 To see this one first writes the solution of eq. (\ref{n1loopRG}) in
the standard ``non-improved'' way as
$x(Q^2/\mu^2, x_0)$, where $x_0\equiv x(Q^2=\mu^2)$, and expands in powers of 
$x_0$. 
The resulting coefficients are polynomials in $\log(Q^2/\mu^2)$ 
(the standard RG logs) and can all be expressed in terms of the 
$\beta$ function coefficients. 
If the $\beta$ function is given by a convergent power series 
(as in all examples we examine here), then the resulting series for
$x(Q^2/\mu^2, x_0)$ has a {\em finite} radius of convergence at any fixed 
$\vert Q^2/\mu^2\vert$, and defines the unique correct analytic
continuation to complex $Q^2$. 
Given the finite convergence radius, there are no singularities
for a fixed $\vert Q^2/\mu^2\vert$ if $x_0$ is small enough. 
Due to asymptotic freedom, small enough $x_0$ at
fixed $\vert Q^2/\mu^2\vert$ corresponds to large $\vert Q^2 \vert$
and therefore $x(Q^2)$ has no singularities for large enough $\vert
Q^2\vert$.

Considering the 2-loop APT spectral function,
\hbox{$\beta_R(t)=-(1/\pi)\,{\rm Im}\{x(-s-i\epsilon)\}$} 
with $x(Q^2)$ given by
eq. (\ref{W_sol_2loop}), we show in fig.~4 the values of $W$ along the
time-like axis, below the cut, for a 2-loop $\beta$
function with $\beta_0=1$ and several different values of $c$.
We identify two categories of lines  in fig.~4:
\begin{description}
\item{(a) }
for $c<-\beta_0$ ($c<-1$ in the figure) $W$ flows to zero in
the infrared. This leads to a non-trivial perturbative infrared
fixed-point: $x(Q^2)\longrightarrow x_{FP}=-1/c$. 
\item{(b) }
for $c>-\beta_0$ (where $c$ can be either positive or negative) 
$\vert W\vert \longrightarrow \infty$ in the infrared, and thus 
$x(Q^2)\longrightarrow 0$. 
\end{description}

When $c<-\beta_0$,
the singularity structure (see fig.~3) is consistent
with causality: the cut is only on the time-like axis. This
guarantees that $x(Q^2)$ obeys the dispersion relation in eqs.
(\ref{beta_R}) and (\ref{DR}). It follows that in this case the APT coupling
coincides with the perturbative coupling: $x_{\APT}(Q^2)=x(Q^2)$. 
On the other hand, when $c>-\beta_0$, the singularity structure is always
inconsistent with causality (see fig.~3). Therefore in this case 
$x_{\APT}(Q^2)$ differs from the perturbative coupling $x(Q^2)$.

Having identified how the relative size of $c$ vs. $\beta_0$ affects the
main features of the APT coupling, we are now ready to consider the actual
2-loop $\beta$ function in QCD 
and examine the dependence of the analytically continued coupling $x(-s)$
on $N_f$, the number of light quark flavors.
In fig.~5 we plot values of $W$ along the time-like axis, below the cut,
for $N_f=0,1,2,\ldots,16$.
As already mentioned, the branch $W_n$ in which the solution
corresponding to the time-like axis resides, depends on
$c/\beta_0$. The table below lists the values of $N_f$, and the
corresponding branch labels $n$. 
\[
\begin{array}{|r|r|r|r|r|r|r|r|r|r|r|r|r|r|r|r|r|r|}
\hline
N_f& 0 & 1 & 2 & 3 & 4 & 5 & 6 & 7 & 8 & 9 & 10 & 11 & 12 & 13 & 14 & 15 & 16\\
\hline
n& 1 & 1 & 1 & 1 & 1 & 1 & 1 & 2 & 25 & -1 & 0 & 0 & 0 & 0 & 0 & 0 & 0 \\
\hline
\end{array} 
\]
For $N_f\geq 10$ we have $c<-\beta_0$, so
the APT coupling coincides with the perturbative one. On the other hand,
for $N_f\leq 9$ we have $c>-\beta_0$, and then we expect the APT solution to be
different. For $N_f=9$, $-\beta_0<c<0$ and therefore a pair of complex
singularities appears, while for $N_f\leq 8$ there is a single singularity of
the space-like axis. The $N_f=8$ line is not shown in fig.~5, since the large
value of $\beta_0/c$ in this case causes the corresponding
Im($W$) to fall within the $W_{25}$ branch, way outside the vertical range 
of the plot.
It is important to note that for the physically relevant case,
$N_f\leq3$, the exact number of flavours has only a small effect on $W$.

Next, we use (\ref{W_sol_2loop}) to calculate $x(-s)$ below the cut
for two representative examples: $N_f=3$ and $N_f=10$. The corresponding
real and imaginary parts of $x(-s)$ are shown in fig.~6 as function of
$t=\ln(s/\Lambda^2)$.  The ultraviolet limit of the real part is the same
in both cases: ${\rm Re}\{x(-s)\}\,\rightarrow\, 0$.
The difference is in the infrared:
for \hbox{$N_f=10$}, \
\hbox{${\rm Re}\{x(-s)\}\,\rightarrow \,x_{FP}=-1/c$}, 
in accordance with (a) above, while for $N_f=3$, 
\hbox{${\rm Re}\{x(-s)\}\,\rightarrow \,0$}, 
in accordance with (b) above.  
The imaginary part of $x(-s)$ vanishes in both the infrared and in the 
ultraviolet limit, 
so that the APT coupling given by (\ref{beta_R}) and (\ref{DR}) is well
defined.

${\rm Im}\{x(-s)\}$ below the cut is always positive-definite 
and thus the APT coupling is a
monotonically decreasing function of $Q^2$, attaining
its maximum value $x_{\APT}(0)$  at $Q^2=0$.
The infrared limit of the APT coupling can be obtained by 
integrating ${\rm Im}\left\{x(-s)\right\}$ over all scales:
\beq
x_{\APT}(0)=\frac{1}{\pi}\int_{-\infty}^{\infty}{\rm Im}\left\{x(-s)\right\} 
\,dt
\label{allscaleint}
\eeq  
This integral can be performed analytically by
changing the integration variable to $W$. Using (\ref{LambertW_def}) and 
(\ref{W_sol_2loop}) we write $dt=(-c/\beta_0)(1+W)/W\,dW$ and obtain: 
\beq
x_{\APT}(0)=\frac{1}{\beta_0}\,\frac{1}{\pi}\,
{\rm Im}\left\{\int_{\rm IR}^{\rm UV}\frac{dW}{W}\right\} 
\label{as_2loops}
\eeq
where the integration contour in the $W$ plane is along the line that
corresponds to the {\em time-like} axis (examples are provided by the
continuous lines in figs. 4 and 5). 

For $c>0$ and for $-\beta_0<c<0$ the time-like axis is always mapped
into a non-principle branch ($n\neq 0$) and then the 
integration contour stretches from ${\rm Re}\{W\}\longrightarrow -\infty$ 
to ${\rm Re}\{W\} \longrightarrow +\infty$, while ${\rm Im}\{W\}$ is
bounded. The integral can then be performed by closing the
integration contour with a semi-circle at infinity. Since
there are no poles inside the closed contour and the integration along
the semi-circle yields $i \pi$, one obtains $x_{\APT}(0)=1/\beta_0$.
    
For $c<-\beta_0$ the relevant branch is $n=0$ and then the contour
starts at $W=0$ in the infrared and reaches 
${\rm Re}\{W\} \longrightarrow +\infty$ in the ultraviolet. In this
case we evaluate the integral directly and then use the
asymptotic behavior of $W_0(z)$ at small and large $\vert z\vert$. The
result is $x_{\APT}(0)=-1/c$.
A posteriori, this should not come as a surprise,
since we already know that whenever the time-like coupling is within
the $W_0$ branch, ($c<-\beta_0$), the APT solution coincides with the 
perturbative coupling which flows to the infrared fixed-point
at $-1/c$. 

We note that the conclusion of refs.~\cite{Analytic_1,Analytic} that the
infrared limit of the 2-loop APT coupling is always $1/\beta_0$ is correct
only for $c>-\beta_0$.

Our results for 2-loop $x_{\APT}(0)$ are summarized in fig.~7, where the
upper plot (continuous line) 
shows the 2-loop APT infrared limit for a $\beta$ function with
$\beta_0=1$ and a span of $c$ values, and the lower plot (continuous
line and crosses) shows the 2-loop
APT infrared limit for QCD with a varying number of flavours. 
For $N_f\leq9$ the APT infrared limit is $1/\beta_0$, while for
$N_f\geq10$ it is $-1/c$.   

In a similar manner we now calculate the infrared APT limit for the
PA-improved three-loop coupling defined from the imaginary part of
$x(-s)$ in eq. (\ref{W_sol_3loop}). Also here we perform the
integral analytically by changing the integration variable to $W$, 
\beq
x_{\APT}(0)=\frac{1}{\beta_0}\,\frac{1}{\pi}\,
{\rm Im}\left\{\int_{\rm IR}^{\rm UV}\frac{W+1}{W+1-c_2/c^2}
\,\frac{dW}{W}\right\}. 
\label{as_3loops}
\eeq
We find the same two scenarios: when the time-like axis
corresponds to a non-principal branch, we close the contour by a
half-circle at infinity. Otherwise, for $n=0$, we evaluate the
integral directly and use asymptotics of $W_0(z)$. 

Depending on the values of $\beta_0$, $c$ and $c_2$, there are four
different possibilities, as summarized below:

\vskip 20pt

\beq
\begin{array}{ll}
\begin{array}{ll}
 &  \\
 &  \\
c>0 &  \\
c<0  & \left\{ \begin{array}{ll}
c>-\beta_0 & \\
c<-\beta_0  & \left\{ \begin{array}{ll}
c_2<c^2 & \\
c_2>c^2 & \\
\end{array}\right.
\end{array}\right.
\end{array}&
\begin{array}{l}
\underline{x_{\APT}(0)}\\
\\
1/\beta_0\\
1/\beta_0\\
(-1/c)/(1-c_2/c^2)\\
1/\beta_0-(1/c+1/\beta_0)/(1-c_2/c^2)
\end{array}
\end{array}
\label{3_loops_APT_0}
\eeq
\vskip 20pt

For $c>-\beta_0$ we again obtain a completely universal infrared
limit: $x_{\APT}(0)$ does not depend at all on $c$ and $c_2$.
The universality breaks down for $c<-\beta_0$, just like in the two-loop case.
 When the singularity structure of the perturbative
coupling is consistent with causality, we expect that it coincides
with the APT coupling, and then $x_{\APT}(0)$ should be just equal to
the solution of $\beta(x)=0$,  which is $x(Q^2=0)=(-1/c)/(1-c_2/c^2)$.
However, contrary to the two-loop case, at three-loops, given
$c<-\beta_0$, it is still
possible that the singularity structure will be inconsistent with
causality: if $c_2>c^2$ the perturbative coupling has 
an extra pole on the space-like axis (see discussion below eq. 
(\ref{scheme_trans}), case (b)). Indeed, in this case the APT infrared
limit is different from the perturbative one.

Our results for the infrared limit of the APT coupling at
3-loops are shown in fig.~7, on top of the 2-loop results. The plot
demonstrates that as long as $c>-\beta_0$ the limit
$x_{\APT}(0)=1/\beta_0$ is universal, i.e. it does not depend on $c$
and $c_2$. Nevertheless, when $c<-\beta_0$ the values of $c$ and $c_2$
are important for the APT infrared limit, as indicated by
(\ref{3_loops_APT_0}).  

In the upper plot, one notes that for a 3-loop case with $c_2>0$ and $c<0$,
$x_{\APT}(0)$ can become large, and even diverge, when $c_2\sim c^2$. 
Then $x_{\APT}$ ceases to be a good expansion parameter.
Another interesting feature of the 3-loop solution is the jump that may
occur in the value of $x_{\APT}(0)$ at $c=-\beta_0$ when $c$ is
varied. Such a jump indeed occurs if $c_2<c^2$ (dashed line in the
upper plot).   

In the lower plot, one observes that 
the APT coupling based on the PA-improved 3-loop
$\beta$ function in $\MSbar$ shows a sharp transition at $N_f\simeq 10$. 
For $N_f\geq 10$ the coupling has a perturbative infrared fixed point
at a rather small value ($x(0)\sim 0.1$), suggesting a reliable purely
perturbative phase. On the other hand, for $N_f\leq 9$ the APT coupling has a
non-perturbative infrared fixed point, with the ``universal'' value of
$1/\beta_0$. It is natural to ask whether the transition
observed at $N_f\simeq 10$ is an indication of the phase
transition expected in QCD as $N_f$ is increased (see \cite{FP} and
refs. therein), or is it an artifact
of considering a truncated $\beta$ function in a special 
renormalization scheme. While we cannot fully answer this
question, it is worthwhile noting that the transition
point $N_f\sim 10$ does not depend on the renormalization scheme, since
it reflects the {\em scheme-independent} condition
$c=-\beta_0$. 
Moreover, the essential qualitative features of the $\MSbar$ results
for $N_f\geq 10$ are likely shared by most acceptable renormalization
schemes:  one observes that in physical schemes the condition $c_2<c^2$ 
is always obeyed for $N_f\geq 10$ (see fig.~1 in \cite{FP}).
Therefore, the perturbative fixed-point of the PA-improved 3-loops
$\beta$ function at $(-1/c)/(1-c_2/c^2)$ is positive and the
perturbative coupling has a causal analyticity structure.  The last
case considered in (\ref{3_loops_APT_0}), i.e. $c<-\beta_0$ and
$c_2>c^2$, is probably never realized in QCD.

\section{Summary and Conclusions}

Using the Lambert W function we achieved a thorough understanding
of the structure of the solutions of the 2-loop and (Pad\'e improved) 3-loop
RG equation for the coupling constant in the complex $Q^2$ plane. 
The main result 
is that the running perturbative coupling in QCD can have a causal analyticity
structure, with a non-trivial infrared fixed point, 
provided $\beta_0$ is small enough and that $\beta_1$ is negative, i.e. in the 
framework of the Banks-Zaks expansion. 
This suggests that a consistent, purely 
perturbative definition of QCD should be possible when the number of flavors
is large enough (barring possible complications due to renormalons and
to the asymptotic nature of the $\beta$ function series and of the Banks-Zaks
series, which have not been addressed here).
On the other hand, for larger $\beta_0$ or for positive $\beta_1$,
unphysical singularities are present in the infrared
region, which are to be removed by non-perturbative power-like effects. 

At the 2-loop level, the causality requirement 
translates into the condition \break
\hbox{$\beta_1<-\beta_0^2$}, i.e. $x_{FP}\equiv -1/c<1/\beta_0$.
 This condition
is also relevant for the PA-improved 3-loop case, for most of the
acceptable renormalization schemes. Taking seriously the
2-loop and 3-loop results, we are lead to conclude that a
purely perturbative phase is realized in QCD with $N_f\geq 10$. This
result is in reasonable agreement with other approaches (see \cite{FP}
and refs. therein).

We have investigated the APT approach, a simple mathematical way 
(which admittedly lacks a physical basis, see the discussion below) 
to implement the necessary power-like effects, starting from perturbation
theory. In this approach, a causal coupling is reconstructed via a dispersion
relation from the time-like discontinuity of the {\em perturbative}
coupling. 
We have shown that  in some cases
the value of the APT infrared fixed point may
depend on higher-order coefficients of the perturbative
$\beta$ function, and is not always given by the one-loop value  $1/\beta_0$, 
which is therefore not ``universal''. 
Departure from the ``universal'' value
appears not only in the above mentioned cases where the perturbative coupling
is by itself already causal and the APT fixed point coincides with the 
perturbative one, but also when unphysical space-like singularities are present
and a (positive) perturbative infrared fixed point does not exist
-- case (b) below eq.~(\ref{scheme_trans}).
Nevertheless, the latter case might be viewed as academic, since 
in QCD it is not realized for any number of flavors.

It is therefore natural to wonder what general conditions
are required to recover the ``universal'' APT value, which clearly has a 
special, remarkable status (in particular, all the ``non-perturbative''
APT fixed point curves end up on the ``universal'' $1/\beta_0$ line in
Fig.(7)).
Sufficient conditions for the ``universal'' APT value were given 
in \cite{dispersive_grunberg}, {\em assuming} a Landau singularity is present
on the space-like axis: essentially, the perturbative coupling was required to 
approach the {\em trivial} infrared fixed point when the scale is
decreased to zero on the {\em space-like} axis below the Landau
singularity\footnote{This condition also allows to 
relate \cite{grunberg-fixedpoint} the ambiguity arising from
integrating over the Landau singularity to the one due to renormalons,
a fact exploited in the proof.}. 
Thus, in order to avoid universality in this case, there must exist
a {\em non-trivial}, but {\em perturbative}, infrared 
fixed point (at unphysical negative or complex values of the
coupling), whose domain of attraction
includes the trajectory going through the Landau singularity.
Indeed, a (negative) perturbative infrared fixed point is present 
in the 3-loop example for $c<0$ and $c_2>c^2$, explaining why the last 
``non-perturbative'' APT fixed point in eq. (\ref{3_loops_APT_0}) differs from
the ``universal'' value. 
Actually, universality may be obtained under more general conditions 
then those stated in \cite{dispersive_grunberg}, 
e.g. in the case $-\beta_0<c<0$   in eq. (\ref{3_loops_APT_0}), where 
the APT value is still the universal one, despite the presence of a non-trivial
perturbative infrared fixed point (which always exists for $c<0$), and 
regardless of its sign (if the latter is negative, then a Landau pole is still
present on the space-like axis).

In all the examples presented here, namely the 2-loop and PA-improved
3-loop $\beta$ functions, universality is obtained whenever the 
perturbative coupling approaches the
{\em trivial} infrared fixed point as $Q^2\longrightarrow 0^-$ on the 
{\em time-like} axis. This has a simple mathematical explanation in
terms of the Lambert W function, as in these cases, the contour in the
$W$ plane stretches from ${\rm Re}\{W\}\longrightarrow -\infty$ 
to ${\rm Re}\{W\} \longrightarrow +\infty$, while ${\rm Im}\{W\}$ is
bounded. The integral can then be performed by closing the
integration contour with a semi-circle at infinity. Since
there are no poles inside the closed contour and the integration along
the semi-circle yields $i \pi$, one obtains $x_{\APT}(0)=1/\beta_0$.
As a result, the details of the contour, which do depend on the
coefficients of the $\beta$ function, are insignificant.

It would be interesting to know whether, for a {\em generic}
higher-order $\beta$ function, a universal $1/\beta_0$ limit of the APT
coupling is obtained whenever the perturbative coupling approaches the
{\em trivial} infrared fixed point on the {\em time-like} axis, giving 
a sufficient condition for universality alternative to that 
of \cite{dispersive_grunberg}. Unfortunately, because of the complicated
phase structure in the complex $Q^2$ plane when complex Landau singularities
are present (see the discussion before eq. (\ref{d_n_neg_c})), it is not clear 
if such a condition could turn out to be also necessary. Note that in the 
present examples, for $-\beta_0<c<0$ there is a 2-phase structure, such that
the space-like coupling approaches the non-trivial perturbative infrared
fixed point, while the time-like coupling approaches the trivial one. On the 
other hand, for $c<-\beta_0<0$  there is only one phase and both couplings
approach the non-trivial fixed point. But, in principle, there could be more 
complicated examples where a third scenario is realised, namely the space-like
coupling approaches the trivial fixed point (thus insuring universality
according to \cite{dispersive_grunberg}), while the time-like coupling
approaches a non-trivial one. 

It is also interesting to note that the PA-improved 3-loop coupling 
probably offers the simplest example (in case $c<0$ and $c_2>c^2$) 
where the standard relation  between renormalons and Landau pole
(see previous footnote)
 might not hold, since the Landau pole trajectory is not in the domain of 
attraction of the trivial infrared fixed point in this example.  
Developing this point is beyond the scope of the present paper.

One might question the physical relevance of the APT coupling. It is 
unlikely, given its perturbative origin, that it exhausts all non-perturbative 
effects in the running coupling, but it may play a useful phenomenological role
(yet to be clarified)  in a more general 
framework \cite{dispersive_grunberg,alekseev}.

\begin{flushleft}
{\large\bf Acknowledgments}
\end{flushleft}
This research was supported in part by the Israel
Science Foundation administered by the Israel Academy of Sciences and
Humanities, by a Grant from the G.I.F., the German-Israeli
Foundation for Scientific Research and Development, by the Charles
Clore doctoral fellowship, and by the EC program `Training and Mobility of 
Researchers', Network `QCD and Particle Structure', contract ERBFMRXCT980194.

\newpage
\begin{figure}[htb]
\begin{center}
\mbox{\kern-0.5cm
\epsfig{file=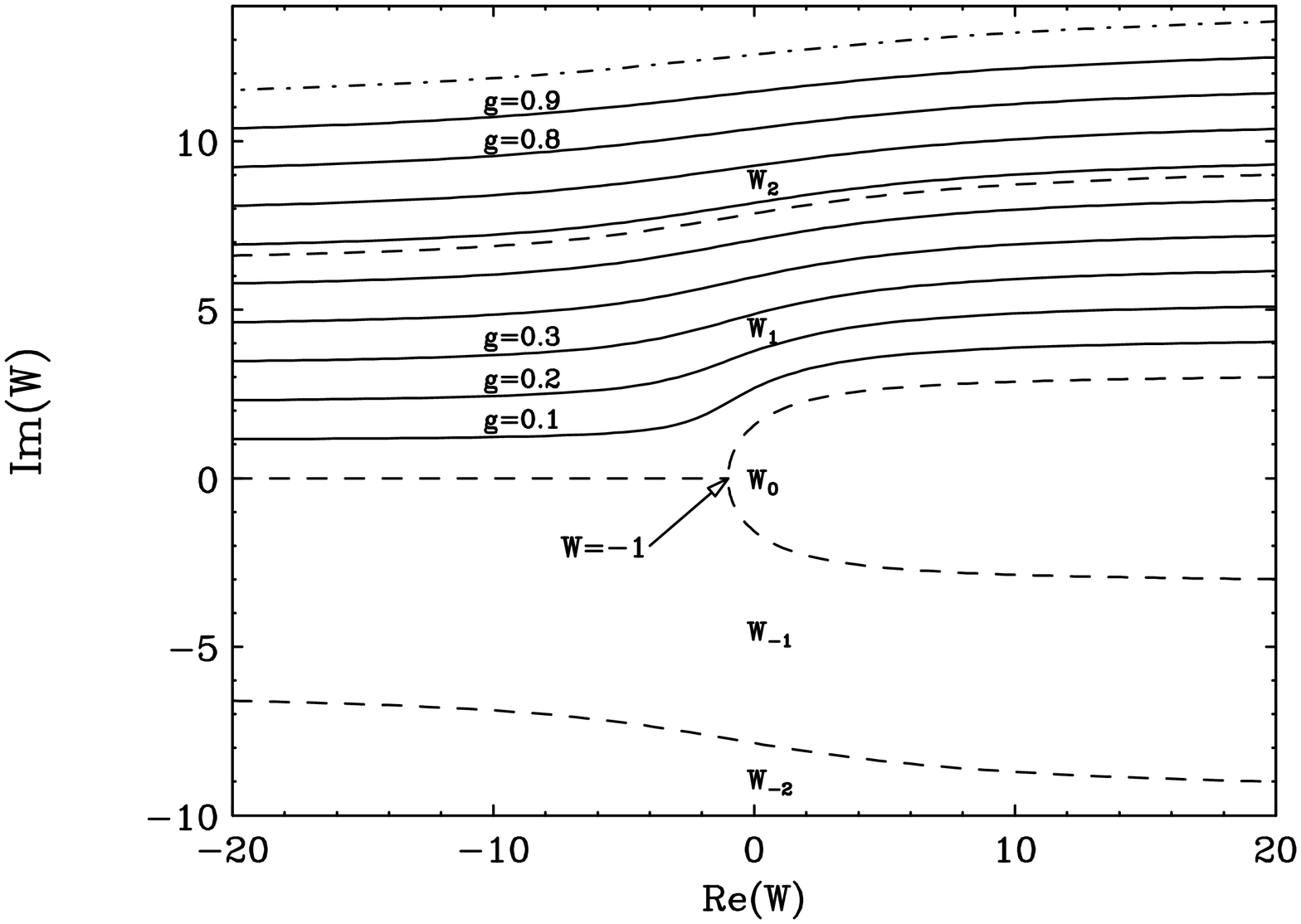,width=10.0truecm,angle=90}
}
\end{center}
\caption{The complex $W$ plane is divided by dashed lines 
to branches of the Lambert W function: $W_0$, $W_{\pm 1}$, $W_{\pm 2}$.
The dashed line that forms the boundary between $W_1$ and $W_{-1}$
represents also the values of $W$ for the space-like coupling at
2-loops for $c>0$ (above the Landau singularity).
The continuous lines
correspond to the values of $W$ for a set of fixed phase rays in the
lower $Q^2$ half-plane for the example of a 2-loop $\beta$ function with 
$\beta_0=1$ and $c=2/7$. Large $\vert Q^2 \vert$ corresponds to the
left side of the lines, while the infrared limit corresponds the right side. 
The phases in the $Q^2$ plane are 
$\phi=-g\pi$, where $g=0.1,0.2,0.3,\ldots, 0.9$. The dot-dashed
line ($g=1$) corresponds to the time-like axis below the cut. The use of
$W_2$ for $\phi<-\phi_1=-2(c/\beta_0)\pi\simeq -0.571\pi$ guarantees
the continuity of $W$.   
 }
\label{w_plane_phase_2}
\end{figure}

\newpage
\begin{figure}[htb]
\begin{center}
\mbox{\kern-0.5cm
\epsfig{file=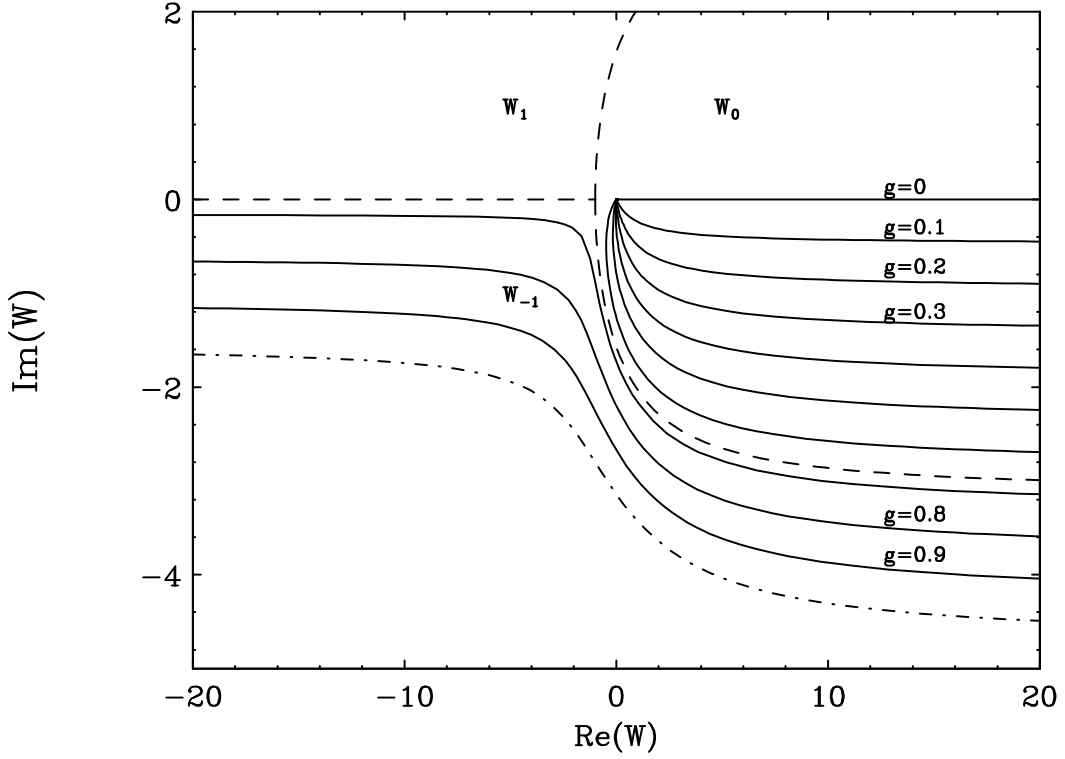,width=10.0truecm,angle=90}
}
\end{center}
\caption{The complex $W$ plane is divided by dashed lines 
to branches of the Lambert W function: $W_0$ and $W_{\pm 1}$.
The $g=0$ line corresponds to the values of $W$ for the space-like
coupling at 2-loops for $c<0$. 
The rest of the continuous lines correspond to the value of $W$ for a set of 
fixed phase rays in the lower $Q^2$ half-plane for the example of a 
2-loop $\beta$ function with $\beta_0=1$ and $c=-2/3$.  
Large $\vert Q^2 \vert$ corresponds to the
right side of the lines.  The phases in the $Q^2$ plane are 
$\phi=-g\pi$, where $g=0.1,0.2,0.3,\ldots, 0.9$. The dot-dashed
line ($g=1$) corresponds to the time-like axis below the cut. The use of
$W_{-1}$ for $\phi<-\phi_0=-\vert c/\beta_0\vert \pi\simeq -0.667\pi$ 
guarantees continuity of $W$ for $\vert Q^2\vert>Q_0^2$ 
(see eq. (\ref{Q0})).  }
\label{w_plane_phase_1}
\end{figure}

\newpage
\begin{figure}[htb]
\begin{center}
\mbox{\kern-0.5cm
\epsfig{file=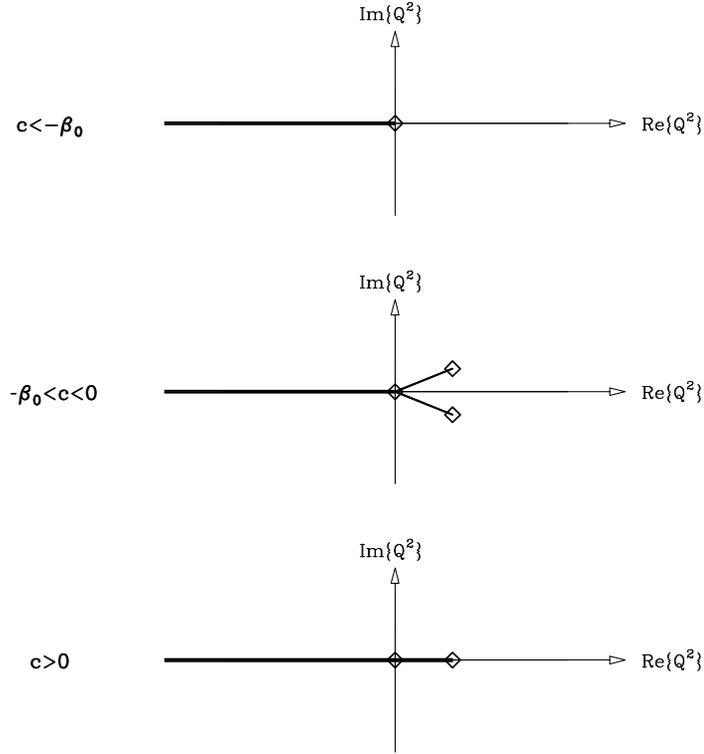,width=10.0truecm,angle=90}
}
\end{center}
\caption{The three possibilities for the singularity structure of the
  2-loop coupling. Branch points are represented by diamonds
  and cuts by bold lines. Only for $c<-\beta_0$ (upper plot)
  perturbative freezing leads to a singularity structure that is
  consistent with causality. For $-\beta_0<c<0$ there is a pair of
  complex conjugate branch points and for $c>0$ there is one
  space-like branch point that violate causality. Note that with 
  the present analytical continuation, the non-causal singularities
  are confined to the infrared region, where they can be removed 
from physical quantities by non-perturbative effects.}
\label{figIII}
\end{figure}

\newpage
\begin{figure}[htb]
\begin{center}
\mbox{\kern-0.5cm
\epsfig{file=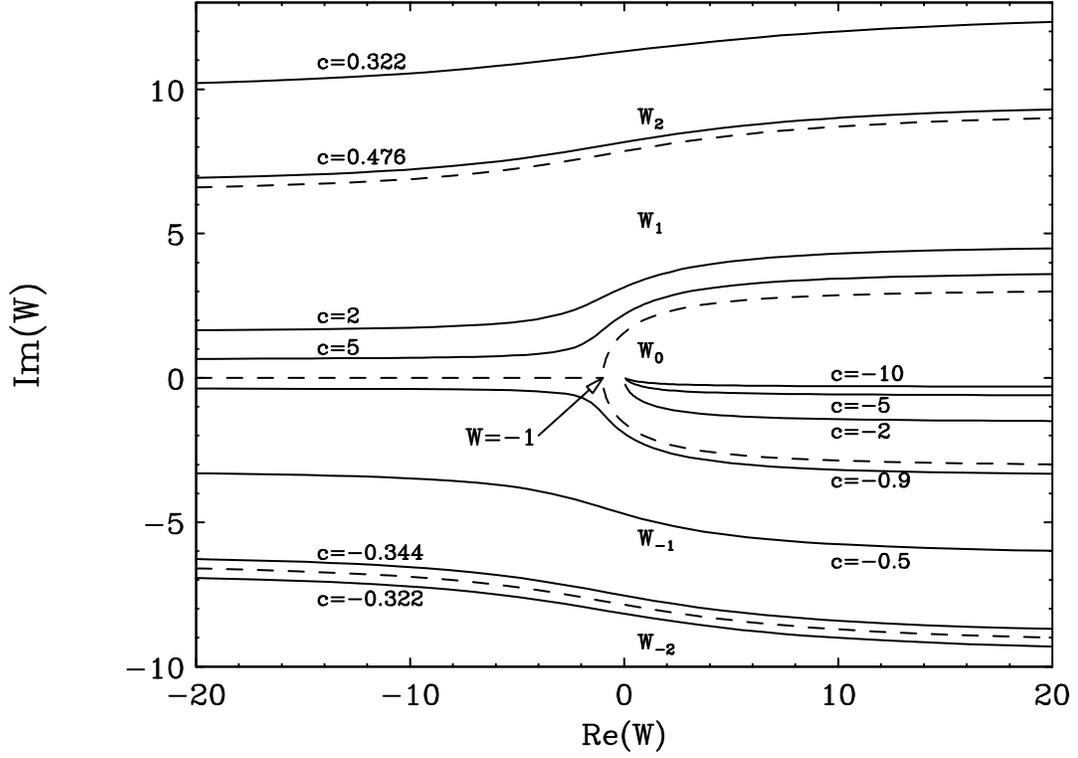,width=10.0truecm,angle=90}
}
\end{center}
\caption{
The various continuous lines show the value of $W$ along the time-like
axis (below the cut), corresponding to the 2-loop coupling for $\beta_0=1$ and
different values of $c$. For the $c>0$ large $\vert Q^2 \vert$ 
corresponds to the left side of the lines, while for $c<0$ large 
$\vert Q^2 \vert$ corresponds to the right side. 
The dashed lines show the division of branches of the Lambert W
function: $W_0$, $W_{\pm 1}$ and $W_{\pm 2}$.
}
\label{w_plane}
\end{figure}

\newpage
\begin{figure}[htb]
\begin{center}
\mbox{\kern-0.5cm
\epsfig{file=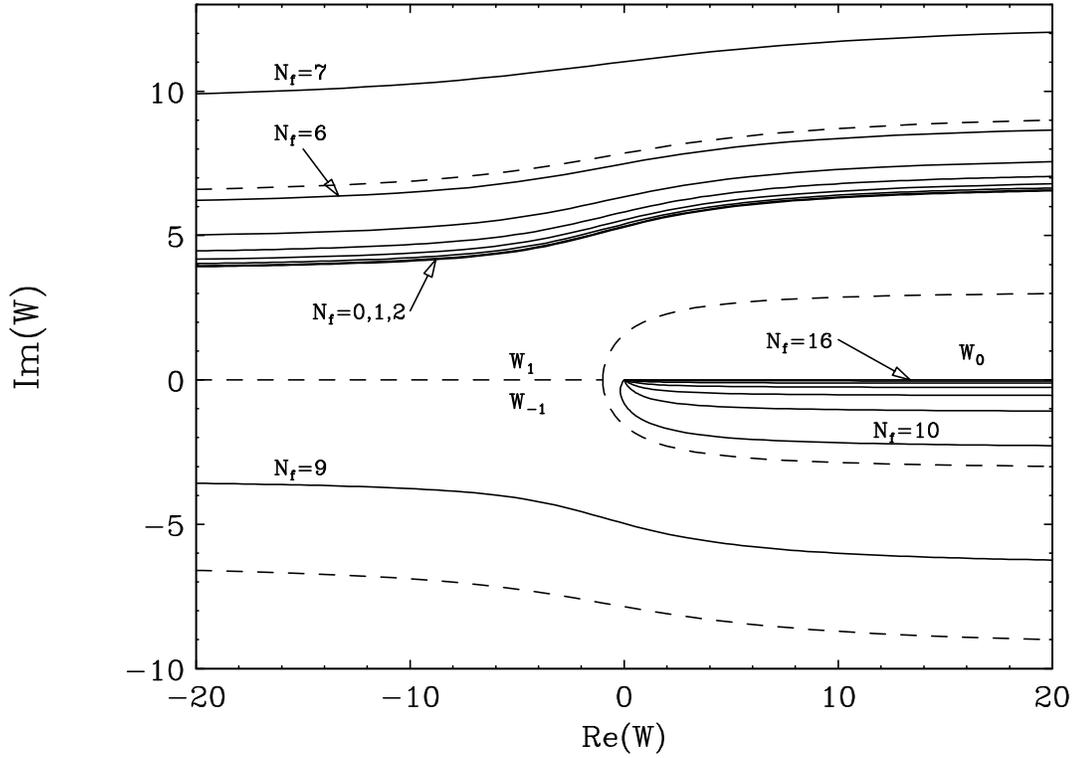,width=10.0truecm,angle=90}
}
\end{center}
\caption{
The various continuous lines show the value of $W$ along the time-like
axis (below the cut), corresponding to the QCD 2-loop coupling for 
different value of $N_f=0,1,2,\ldots, 16$.
The dashed lines show the division of branches of the Lambert W 
function: $W_0$ and $W_{\pm 1}$. 
}
\label{w_plane_Nf}
\end{figure}

\newpage
\begin{figure}[htb]
\begin{center}
\mbox{\kern-0.5cm
\epsfig{file=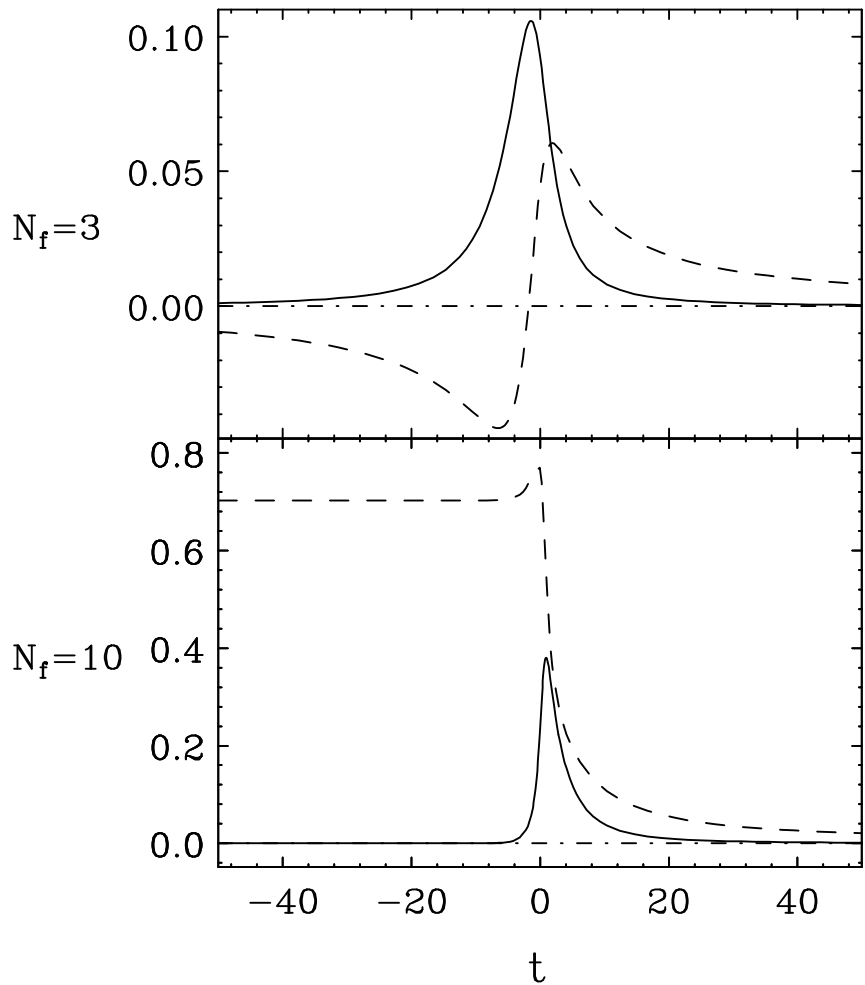,width=13.0truecm,angle=90}
}
\end{center}
\caption{The real (dashed line) and imaginary (solid line) parts
  of the analytically continued coupling on the time-like axis, 
  $x(-s)$, for the case of a 
  2-loop $\beta$ function in QCD with $N_f=3$ ($\beta_0=9/4$ and
  $c=16/9$) in the upper box, and 
  $N_f=10$ ($\beta_0=13/12$ and $c=-37/26$) in the lower box. 
  For $N_f=10$ perturbative freezing leads to a well defined infrared limit 
  for the perturbative coupling: $x^{FP}=-1/c$, while for $N_f=3$ the
  ``perturbative infrared limit'' is zero. 
 }
\label{time_like}
\end{figure}

\newpage
\begin{figure}[htb]
\begin{center}
\mbox{\kern-0.5cm
\epsfig{file=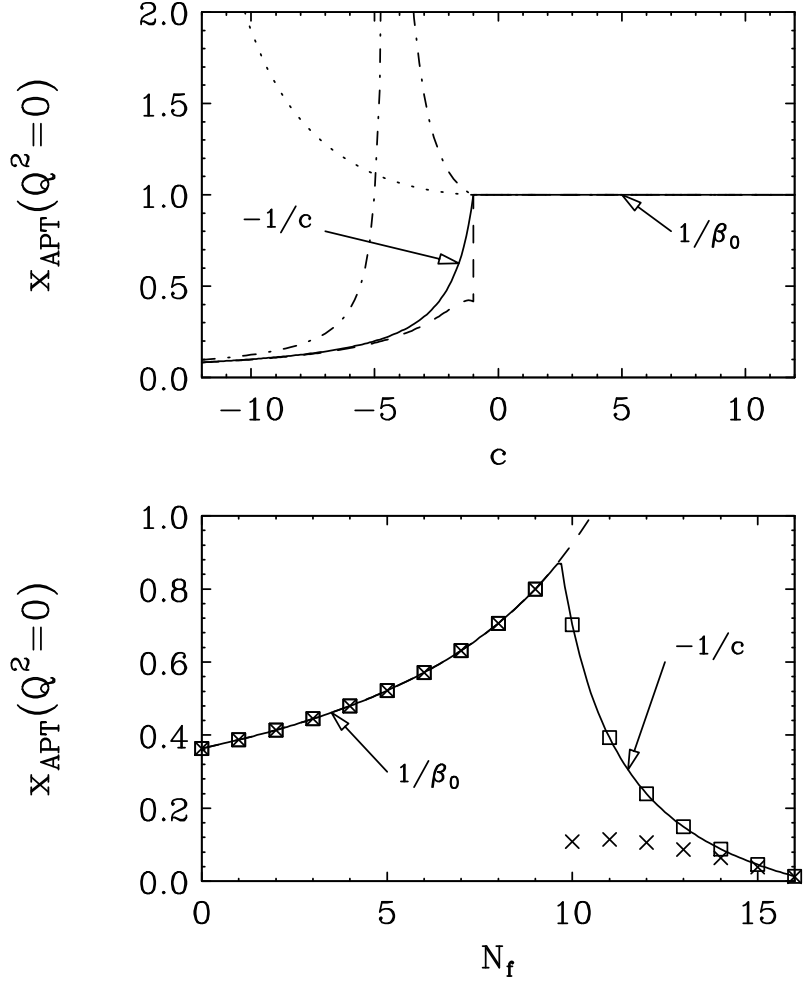,width=13.0truecm,angle=90}
}
\end{center}
\caption{
The upper box shows the infrared limit of the APT solution for a
  hypothetical PA-improved 3-loop $\beta$ function with $\beta_0=1$, a span of
  values for $c$, $-12\,<\,c\,<\,12$ and various values for $c_2$. 
  Continuous line: $c_2=0$ (the 2-loop case), dashed line:
  $c_2=-1.4$, dot-dashed line: $c_2=20$, dotted line: $c_2=200$. 
The lower box shows $x_{\APT}(0)$ in QCD with $0\leq N_f\leq 16$. 
Squares and crosses correspond to a 2-loop and a PA-improved 3-loop  
$\MSbar$ $\beta$ function, respectively. 
For $N_f\leq 9$, $x_{\APT}(0)$ is $1/\beta_0$ in both cases.}
\label{x_fp}
\end{figure}

\end{document}